
\magnification=1200
\baselineskip=16pt
{}.
\vskip 2cm
\centerline{\bf Critical Behavior of Hierarchical Ising Models}
\vskip 2cm
\centerline{Ferenc Igl\'oi$^{\dag,\ddag}$, P\'eter
Lajk\'o$^{\ddag}$ and Ferenc Szalma$^{\ddag}$}
\vskip 1cm
\centerline{$^{\dag}$Research Institute for Solid State Physics}
\centerline{H-1525 Budapest, P.O.Box 49, Hungary}
\smallskip
\centerline{and}
\smallskip
\centerline{$^{\ddag}$Institute for Theoretical Physics, Szeged University}
\centerline{H-6720 Szeged, Aradi V. tere 1, Hungary}
\vskip 2cm
{\bf Abstract:}
We consider the critical behavior of two-dimensional layered Ising models
where the exchange couplings
between neighboring layers follow hierarchical sequences. The perturbation
caused by
the non-periodicity could be irrelevant, relevant or marginal. For marginal
sequences we have performed a detailed study,
which involved analytical and numerical calculations of different surface
and bulk critical
quantities in the two-dimensional classical as well as in the one-dimensional
quantum version of the model. The critical exponents are found to vary
continuously with the strength of the modulation, while close to the critical
point the system is
essentially anisotropic: the correlation length is diverging with different
exponents along and perpendicular to the layers.

\vskip 1cm
PACS-numbers: 05.50.+q, 64.60.Cn, 64.60.Fr
\vfill
\eject
{\bf I. Introduction}
\vskip 1cm
Since the discovery of quasicrystals[1] there is a growing interest to
understand
their structure and physical properties. Theoretically it is a challenging
problem to understand the properties of phase transitions in these
quasiperiodic or more generally non-periodic structures. Since a non-periodic
system shares some aspects with a system with quenched disorder, from these
studies one hopes to get a better understanding about the critical behavior of
random systems, too.

Early studies on this field were numerical investigations on specific problems
(Ising model[2-4], percolation[5,6], random walks[7] etc.) on two- and
three-dimensional
quasi\-periodic lattices, whereas analytical results were obtained on layered
two-dimensional Ising models with one-dimensional aperiodicity[8,9] or on the
corresponding Ising quantum chain[10-14]. The nature of phase transitions is
better
understood, since Luck[15] generalized the Harris criterion[16] for
non-periodic
systems. In this relevance-irrelevance criterion the strength of fluctuations
in the couplings in a scale of the correlation length is of primary importance.
The perturbation caused by the
non-periodicity is irrelevant (relevant) if the local energy fluctuations are
smaller (greater) than the corresponding thermal energy. Theoretically most
interesting is the borderline case, when the thermal and fluctuating energy
contributions are in the same order of magnitude, i.e. the perturbation is
marginal. In this case - according to exact results on two-dimensional layered
Ising models[17-20] - the critical behavior is non-universal, the critical
exponents
are continuous functions of the stregth of the modulation. Furthermore,
close to the critical
point the system becomes essentially anisotropic[20], i.e. the correlation
length
is diverging with different exponents along and perpendicular to the layers.

The Luck criterion is obtained in the frame of a linear stability analysis
therefore its validity is restricted to such non-periodic systems where the
perturbation
in the local couplings is small. Thus the criterion is valid for
quasiperiodic lattices and for such one-dimensional aperiodic sequences wich
are generated by substitutional rules. On the other hand the criterion is no
longer valid in its original form, if the modulation
of the couplings follows some hiererchical sequence, such that certain
couplings
could become arbitrarily large or small. This type of hierarchical
sequence[21]
was introduced first by Huberman and Kerszberg[22] and generalized
by others[23,24]
to study anomalous diffusion and the properties of the spectrum of
the Hamiltonian[25] in one-dimension.

As far as the critical properties of Ising quantum chains with a hierarchical
structure in the couplings are concerned two conflicting results exist. From
a study of the
low-lying
excitations of the system Ceccatto[26] has drawn the conclusion, that the
perturbation is irrelevant, if the hierarchical parameter $r$ is smaller than
some critical value $r_c$. Whereas for $r>r_c$ the perturbation is
relevant, the Ising-type critical point of the homogeneous system is
washed out by the perturbation and the critical behavior of the system
is similar to that of the McCoy-Wu model[8], i.e. to a layered Ising model
with quenched
randomness. In contrary to this results Lin and Goda[27] obtained
continuously varying surface magnetisation exponent of the
hierarchical
quantum Ising model, thus according to this study the perturbation
caused by this non-priodicity
is marginal in the whole range of the parameter $r$. Similar
conclusion is drawn from renormalization group and Monte Carlo
simulation studies on the two-dimensional Ising model with layered
hierarchical couplings[28].

Our aim in the present paper is to clarify this controversary and to present a
comprehensive picture about the critical behavior of hierarchical Ising
models. For this purpose we consider general hierarchical sequences and
investigate the condition for relevance (irrelevance) of the
perturbation. For marginal
hierarchical modulation we perform a detailed study, which includes analytical
and numerical calculations of different bulk and surface critical quantities
both in the two-dimensional classical and in the one-dimensional quantum
version
of the model. Finally, the results are discussed in the frame of a general
scaling theory, which is then compared to the analogous one of aperiodic
systems.
\bigskip
{\bf II. The model}
\vskip 1cm
We consider the Ising model on the square lattice with different
layered structures: the system is translationally invariant either
along the columns (Fig 1a) or along the diagonals (Fig 1b).
In the first case the interaction
along the layers $K_1$ is constant, whereas it is modulated
in the other direction and given as $K_2(k)$ in units of $1/k_B T$ between
neighboring layers at $k$ and $k+1$. In the extreme anisotropic limit
$K_1 \to \infty$, $K_2(k) \to 0$ the transfer matrix of the problem involves
the Hamiltonian of a quantum Ising chain:
$$H=-{1 \over 2} \sum_k [\sigma_k^z+ \lambda_k \sigma_k^x \sigma_k^x]
\eqno(1)$$
where $\sigma_k^x$, $\sigma_k^z$ are Pauli matrices at site $k$ and

$\lambda_k=-2K_2(k)/{\rm ln}({\rm tanh} K_1)$.

In the second situation the $K_d(k)$ diagonal couplings are the
same within one column and the quantities
$$Y_k={\rm sinh}[2K_d(k)] \eqno(2)$$
are assumed to vary in a hierarchical way.

The hierarchical sequences we use in this paper were first
introduced in economical problems[21] and later applied to study the
so called hyperdiffusion process. For an integer number
$m$
the sequence $a_1,a_2,\dots$ is defined as:
$$a_k=a r^{f_k} \eqno(3)$$
where $r$ is the ratio of the sequence, $a$ is some reference value
and the $f_k$-s are natural numbers satisfying the relation:
$$k=m^{f_k}(ml+\mu)~~,~~l=0,1,\dots ~~,~~\mu=1,2,\dots,m-1 \eqno(4)$$
As pointed out in Ref[29] it is possible to generalize the sequence by
modifying
eq(3) as $a_k=a r^{g(f_k)}$, where $g(x)$ is some analytical
function. Here we shall consider power functions:
$g(x)=x^{\omega}$, $\omega>0$, thus the original sequence in eq(3) corresponds
to $\omega=1$.

For the layered Ising models we assume the hierarchical variation
in the couplings $\lambda_k$ and in the
parameters $Y_k$, respectively.
\bigskip
{\bf III. Surface magnetisation}
\vskip 1cm
For the two-dimensional Ising model the surface magnetisation $m_s$ is
the simplest order parameter, which can be most easily determined in the
extreme anisotropic limit eq(1) through the formula:
$$m_s=\left(1+\sum_{j=1}^{\infty} \prod_{k=1}^j \lambda_k^{-2} \right)^{-1/2}
\eqno(5)$$
For the general hierarchical sequence containing the exponent
$\omega$ the surface
magnetisation is rewritten in the form:
$$m_s=\left[ S(\lambda ,r)\right]^{-1/2}~~,
{}~S(\lambda ,r)=\sum_{j=0}^{\infty}\lambda^{-2j} r^{-2n_j}
{}~,~~n_j=\sum_{k=1}^j (f_k)^{\omega}~~,~~n_0=0 \eqno(6)$$
The critical coupling $\lambda_c$ is such that[30]
$$\lim_{j \to \infty} {1 \over j} \sum_{k=1}^j \ln \lambda_k=0
\eqno(7)$$
and related to the hierarchical parameter $r$ as
$$\lambda_c=r^{-\delta(\omega,m)} \eqno(8)$$
where $\delta(\omega,m)=\lim_{j \to \infty} n_j/m^j$.
As shown in the Appendix:
$$\delta(\omega,m)=\left(1-{1\over m}\right) \sum_{j=1}^{\infty}
{j^{\omega} \over m^j} \eqno(9)$$
which can be expressed in closed form for integer $\omega$s.
For $\omega=1,2$ and 3 it is given as:
$$\delta(1,m)={1 \over m-1}~~,~~\delta(2,m)={m+1 \over (m-1)^2}~~,
{}~\delta(3,m)={m^2+4m+1 \over (m-1)^3} \eqno(10)$$

To calculate the surface magnetisation we do separately for $\omega=1$
and $\omega\ne 1$. In the first case, for $\omega=1$ one can verify that
$f_{mp}=f_p+1$ and $f_{mp+\mu}=0~,~\mu=1,2,\dots,m-1$, from which the
relations
$n_{mp}=n_p+p~~,~~n_{mp+\mu}=n_{mp}$ follow. Then deviding the sum for
$S(\lambda,r)$ in eq(6) into $m$ parts:
$$S(\lambda,r)=\sum_{p=0}^{\infty}\lambda^{-2mp} r^{-2n_{mp}}+
\sum_{\mu=1}^{m-1}
\sum_{p=0}^{\infty}\lambda^{-2(mp+\mu)} r^{-2n_{mp+\mu}} \eqno(11)$$
and using the relations between the $n_j$-s one gets the following equation:
$$S(\lambda,r)=S(\lambda^m r,r){1-\lambda^{-2m} \over 1-\lambda^{-2}}
\eqno(12)$$
{}From this expression the surface magnetisation exponent $\beta_s$ defined by
$m_s(t)\sim (t)^{\beta_s}$ as $t=1-(\lambda_c/\lambda)^2 \to 0^+$ can be
evaluated as in Ref[18]. According to eq(6) close to the critical point
$S(t)\sim (t)^{-2\beta_s}$,
and the critical exponent corresponding to eq(12) is given by:
$$\beta_s={\ln \left[{1-\lambda_c^{-2m} \over 1-\lambda_c^{-2}}\right] \over
2 \ln m} \eqno(13)$$
In the special case $m=2$ one recovers the result in Ref[27].
According to eq(13) for $\omega=1$ the critical behavior of the
hierarchical Ising model is non-universal, since $\beta_s$ is a
continuous function of the ratio $r$. This functional
dependence is shown on Fig 2. for several values of the parameter
$m$.

Next we turn to discuss the situation for $\omega\ne 1$, in which
case the surface magnetisation is studied on large finite
sequences. Let us consider the system of length $j=m^N-1\equiv L-1$, i.e.
after $N$ generations, and define in analogy with eq(6):
$$S_N(\lambda,r)=1+\sum_{j=1}^{m^N-1} \prod_{k=1}^j \lambda_k^{-2}=
\sum_{j=0}^{m^N-1}\lambda^{-2j} r^{-2n_j} \eqno(14)$$
This quantity satisfies the relation:
$$S_{N+1}(\lambda,r)={1-X^{2m} \over 1-X^2} S_N(\lambda,r)~~,~~
X=\lambda^{-L}r^{-n_L} \eqno(15)$$
As shown in the Appendix in the large $L$ limit the finite size
corrections to the criticality condition in eq(8) are logarithmic:
$$n_{L}=\delta(\omega,m) L-{\omega \over m-1}
N^{\omega-1}+O\left(N^{\omega-2}\right)  \eqno(16)$$
thus the parameter $X$ in eq(15) in leading order is given as
$X=(\lambda_c/\lambda)^L r^{[\omega/(m-1)] N^{\omega-1}}$ and
at the critical point one gets the relation:
$$S_{N+1}(\lambda_c)={1-Z^{2m} \over 1-Z^2} S_N(\lambda_c)~~,~~
Z=r^{[\omega/(m-1)] N^{\omega-1}} \eqno(17)$$
where $r$ is related to $\lambda_c$ through eq(8).

For large $N$ the parameter $Z$ in eq(17) scales differently for
$\omega=1$, $\omega<1$ and $\omega>1$, respectively, and the
corresponding functional form of $S_N(\lambda_c)$ is also
different in the three cases. In the borderline case $\omega=1$,
which is studied already before, $S_N(\lambda_c)$ has a power law
dependence on the size of the system:
$$S_N(\lambda_c) \sim \left(L \right)^{2x_s}~~,~~\omega=1
\eqno(18)$$
where according to finite size scaling[31] $x_s$ is the scaling
dimension of surface spins. Using the value of $\delta(1,m)$ in
eq(10) one gets from eqs(17) and (18) that $x_s$ corresponds to
$\beta_s$ in eq(13), thus the scaling relation $\beta_s=\nu x_s$ is
satisfied, since the correlation length critical exponent for the
two-dimensional Ising model is $\nu=1$.

By finite size scaling one may investigate the surface critical
behavior of the system at the right end of the chain taking
$L=m^N$ spins. \footnote{$^{\dag}$}{If the chain consists of $m^N-1$ spins as
before,
then it is symmetric to its center and the surface magnetisation is
the same at both ends.}
Denoting the inverse square of the finite size surface
magnetisation as $\bar S_N(\lambda,r)$ one can write the simple
relation:
$$\bar S_N(\lambda,r)=1+\lambda^{-2} r^{-2N} S_N(\lambda,r)
\eqno(19)$$
from which $\bar x_s$ the scaling dimension of surface spins on the
right end
of the chain is given by:
$$\bar x_s=x_s-{{\rm ln} r \over {\rm ln} m}={\ln \left[{1-\lambda_c^{2m}
\over 1-\lambda_c^{2}}\right] \over
2 \ln m} \eqno(20)$$

thus $\bar x_s(\lambda_c)=x_s(\lambda_c^{-1})$.

For $0<\omega<1~~Z$ in eq(17) goes to zero, and the relation in
eq(17) in leading order reads as
$$S_{N+1}(\lambda_c)=m(1+\omega N^{\omega-1} \ln r) S_N(\lambda_c)
{}~,~~0<\omega<1 \eqno(21)$$
which is solved by the function $S_N(\lambda_c) \sim m^N
r^{N^{\omega}}$. Now using from finite size scaling theory[31] that
$m^N\simeq L \sim |t|^{-\nu}$ one obtains with $\nu=1$ for the
temperature dependence of the surface magnetisation:
$$m_s(t) \sim t^{1/2}
r^{-\left(|\log t|/log m\right)^{\omega}/2}~~,~~0<\omega<1 \eqno(22)$$
Thus the surface magnetisation in this case vanishes with the same
exponent $\beta_s=1/2$ as the homogeneous model with $r=1$, however
there is a logarithmic correction. Consequently the hierarchical
perturbation in the copulings for $0<\omega<1$ is {\it marginally
irrelevant}.

The behavior of the system is completely different for
$\omega>1$. In this case one has to study separately $r>1$, when
$Z$ in eq(17) goes to infinity and $r<1$, when $Z$ goes to zero.
In the first case the asymptotic relation
$$S_{N+1}(\lambda_c)=r^{2\omega N^{\omega-1}}S_{N}(\lambda_c)
\eqno(23)$$
is satisfied by the function
$$S_{N}(\lambda_c)\sim r^{2 N^{\omega}} \eqno(24)$$
Therefore the coupling dependence of the surface magnetisation is
anomalous:
$$m_s(t)\sim r^{-\left(|\log t|/\log
m\right)^{\omega}}~~,~~\omega>1~~,~~r>1 \eqno(25)$$
it vanishes faster than any power as $\lambda \to \lambda_c$.

On the other hand for $r<1$ (and $\omega>1$) $Z$ goes to one, and
the asymptotic relation
$$S_{N+1}(\lambda_c)=\left( 1+r^{{2 \omega \over m-1} N^{\omega-1}}
\right) S_N(\lambda_c) \eqno(26)$$
implies $\lim_{N \to \infty} S_N(\lambda_c) < \infty$, since the
product $\prod_{N=N_0}^{\infty}\left( 1+r^{{2 \omega \over m-1}
N^{\omega-1}} \right)$ is convergent. Consequently the surface
magnetisation stays finite at the critical point and the phase
transition at the surface is of first order. According to eq(26)
the surface magnetisation approaches its limiting value $m_s(0)$ in
an anomalous way:
$$m_s(t)-m_s(0) \sim r^{|\log t /\log m|^{\omega-1}
{2\omega/(m-1)}}
\eqno(27)$$
where again the correspondence $L \sim t^{-1}$ was used. The
surface magnetisation for $r=.92$ and different values of $\omega
\ge 1$ are shown on Fig 3.

Next we turn to present results about the critical
behavior of the model at the (1,1) surface, when the non-periodicity
in the couplings is given in the diagonal direction (Fig 1b). The
criticality condition now reads as[32]:
$$\lim_{j \to \infty} {1 \over j} \sum_{k=1}^j \ln Y_k=0 \eqno(28)$$
where the parameter $Y_k={\rm sinh}[2K_d(k)]$ plays a similar role as
$\lambda_k$
for the quantum Ising model in eq(7). To make use further this
analogy we assign the hierarchical modulation in eq(3) to the
generalized $Y_k$ parameters. As a consequence the critical value
$Y_c$ and the ratio of the sequence $r$ are related as in eq(8) for
the quantum model.

To study the magnetisation on the (1,1) surface we use a numerical
procedure by Hilhorst and van Leeuwen (HvL), which is originally
developed for the triangular lattice and based on a repeated use of
the star-triangle transformation[33]. From the square lattice on Fig
1b one can obtain the triangular lattice by connecting vertically
next-nearest neighbors with non-vanishing couplings. The numerical
procedure also works for this triangular lattice,
however, we shall only study the problem on the square lattice. Details
of the
method together with results on the
surface critical behavior of different type of
non-periodic triangular lattice Ising models will be presented
elsewhere[34].

The HvL method enables one to obtain very accurate numerical
estimates on the magnetisation at the (1,1) surface as well as to
study the decay of surface correlations. The obtained results on the
surface critical behavior for marginally irrelevant ($\omega<1$)
and marginally relevant ($\omega>1$) perturbations are
qualitatively the same as we found for the corresponding quantum
chain. In the truely marginal case - $\omega=1$ - the results even
quantitatively agree with those obtained on the (1,0) surface.
We are going to discuss these results in more detail.

First investigating the surface magnetisation one can see that in
the limit $r \to 0~~~m_s(T)=1$ for $T \le T_c$ and $m_s(T)=0$ for
$T>T_c$, i.e. it stays finite at the critical point.
This anomalous behavior is due to the topology
of the lattice in the diagonal direction. As $r\to 0$ in the first
layer (and also in any even layers) $Y_1=Y_c=1/r$, which together
with $K_d(1)$ goes to infinity, resulting an ordered surface layer.
For $r>0$ according to numerical results the surface magnetisation
vanishes at the critical point. For large number of iterations $I$
the critical surface magnetisation goes to zero as a power:
$m_s(t=0)\sim I^{-\gamma}$, where the exponent $\gamma > 0$ is
related to the decay exponent $\eta_{\parallel}$ in eq(29) as
$\gamma=\eta_{\parallel}/4$.

The surface magnetisation exponent
$\beta_s$ determined from the relation
$m_s(t,r) \sim t^{\beta_s}$ is found to
be the same
function of $r$ as that at the (1,0) surface in eq(13). Similar
conclusion holds for the magnetisation exponent on the right end of
the chain. We obtained
$\bar{x}_s(Y_c)=
x_s(Y_c^{-1})$ in close analogy with the results on the
(1,0) surface. Numerical estimates on the surface magnetisation
exponents are drawn on Fig 2.
We note that the accuracy of the calculation is limited by the fact
that the error in the magnetisation decreases very slowly with the
number of iterations $I$ as $I^{-1/2}$.

We have also investigated the decay of critical surface spin
correlations. The decay is given as a power:
$$G_s(r_{\parallel},t=0) \sim r_{\parallel}^{-\eta_{\parallel}}
\eqno(29)$$
and according to the numerical estimates the critical exponent
$\eta_{\parallel}$ can be described with high accuracy by the
formula:
$$\eta_{\parallel}={2 x_s \over x_s+\bar x_s}~~,~~
\bar \eta_{\parallel}={2 \bar x_s \over x_s+\bar x_s} \eqno(30)$$
both for $r<1$ and $r>1$. In eq(30) $\bar \eta_{\parallel}$ is the
decay exponent on the right surface of the system.

The relations in eq(30) are in conflict with ordinary scaling, which
would imply $\eta_{\parallel}=2 x_s$. They could be explained,
however, within the frame of anisotropic scaling theory[35]. Then the
critical spin-spin correlation function at the left side of the
system is expected to behave as:
$$G_s(r_{\parallel},t)=b^{-2x_s}G_s(r_{\parallel}/b^z,b^{1/\nu}t)
\eqno(31)$$
when lengths perpendicular to the layers transform as $r_{\perp}
\to r_{\perp}/b$, whereas scaling along the layers involves the
dynamical exponent $z$: $r_{\parallel} \to r_{\parallel}/b^z$. As a
consequence the correlation length close to the critical point
diverges with different exponents in the two directions:
$\xi_{\perp} \sim t^{-\nu}$ and $\xi_{\parallel} \sim t^{-\nu z}$.
According to eq(31) the decay exponent is given by
$\nu_{\parallel}={2 x_s/z}$, thus the numerical results in eq(30)
imply that the dynamical exponent is expressed by the sum of the
two surface magnetisation exponents:
$$z=x_s+\bar{x_s}={\ln\left[{\lambda_c^{-m}-\lambda_c^{m} \over
\lambda_c^{-1}-\lambda_c}\right] \over \ln m} \eqno(32)$$
In the following we check the validity of the anisotropic scaling
hypothesis by calculating other critical quantities.
\vskip 1cm
{\bf IV. Other critical quantities}
\bigskip
A direct evidence for anisotropic scaling can be obtained from the
behavior of the correlation length. At the critical point the
perpendicular correlation length on a strip of width $L$ is
restricted to $\xi_{\perp} \sim L$, thus $\xi_{\parallel}\sim
\xi_{\perp}^z \sim L^z$. In the extreme anisotropic limit
$\xi_{\parallel}$ is given by the inverse gap of the critical
Hamiltonian in eq(1), thus one expects for low laying states:
$$E_i-E_0 \sim L^{-z} \eqno(33)$$
To check this relation we have numerically studied the excitation
spectrum of the Hamiltonian in eq(1) at the critical point. Using
standard methods[36] a quadratic fermion Hamiltonian is obtained via
the Jordan-Wigner transformation, which is then diagonalized by a
Bogoliubov transformation. Using free boundary conditions we
calculated the energy of the first fermion modes on finite systems
of size $L=m^N-1$ ($m=2$, $N=2,\dots,16$; $m=3$, $N=2,\dots,10$).
Their size dependence is found to be very accurately described by the relation
in eq(33). The $z$-exponents obtained through sequence extrapolation
methods[37] coincide with the values in eq(32) up to $4-5$ digits.

Even more accurate estimates can be obtained if the energy of
fermion modes are calculated in a
second order perturbation expansion[15]. At the critical point their
size dependence follows from the relation:
$$\Lambda(L) \sim \left[\sum_{j=1}^L\sum_{k=1}^L \lambda_{j+1}^2
\lambda_{j+2}^2 \dots \lambda_{j+k}^2 \right]^{-1/2} \eqno(34)$$
Evaluating eq(34) up to $L=2^{24}$ and $L=3^{15}$ the accuracy of
the estimates on the $z$-exponent has been increased up to 7-8
digits.

Next we turn to study the behavior of the specific heat on finite
systems. According to anisotropic scaling the singular part of the
bulk free energy density in a finite system of size $L$ behaves
as[35]:
$$f(t,L)=b^{-(1+z)}f(b^{1/\nu} t,L/b) \eqno(35)$$
This expression is in accord with the finite size scaling behavior
of the singular part of the critical bulk energy density $\epsilon (L) \sim
L^{-x_e}$, since according to our numerical estimates:
$$x_e=z \eqno(36)$$
{}From eq(35) the finite size dependence of the specific heat at the
critical point is given by $C(t=0,L) \sim L^{-\alpha/\nu}$, where
the specific heat exponent
$$\alpha=1-z \eqno(37)$$
is negative for $r \ne 1$. As shown in Table 1 this relation is also
satisfied.

Finally, we report on our results on the scaling behavior of the
surface energy density at the critical point: $\epsilon_s(L) \sim
L^{-x_{es}}$. On the left surface we obtain:
$$x_{es}=z+2 x_s \eqno(38)$$
and a similar expression is valid on the right surface.
\vfill
\eject
{\bf V. Discussion}
\bigskip
In this paper the effect of a layered hierarchical structure on the
critical properties of the two-dimensional Ising model is studied
by analytical and accurate numerical methods. The perturbation is
found to be marginal for $\omega=1$, which corresponds to the
original Huberman-Kerszberg series. In this case the critical
exponents are continuous functions of the hierarchical parameter
$r$, for the whole range of $0<r<\infty$. These results are
in accordance with the findings of Lin and Goda[27] on the surface
magnetisation in the $m=2$ model, however they are in conflict with
the results of Ceccatto[26]. The failure of Ceccatto's calculation
is due to the fact that his perturbational method is only valid for
irrelevant inhomogeneities and can not be used for the hierarchical
sequence, which is a marginal
perturbation. Our
conclusions are
also consistent with the RG and MC simulation results of Stella et al[28] on
a two dimensional hierarchically layered Ising model.

The nature of the hierarchical perturbation is found to be
different for $\omega<1$ and $\omega>1$. In the first case the
perturbation is irrelevant, however there are logarithmic scaling
corrections. In the other regime, for $\omega>1$, the perturbation
is marginally relevant and the surface magnetisation behaves
anomalously at the critical point. In the following we present a
relevance-irrelevance criterion, which explains the above results.

This criterion is actually a modification of the Harris
criterion[16]
for random systems, which is then generalized by Luck[15] to aperiodic
systems. In this criterion one compares the energy $E_f$ due to
fluctuations in the couplings with the thermal excess energy $E_t$.
If the couplings $J_k$ follow a one-dimensional aperiodicity and
their average is $\bar J$ the fluctuating energy in a domain of
size of the bulk correlation length $L\sim \xi$ behaves asymptotically as:
$$E_{fl}(L)=\sum_{k=1}^L (J_k-\bar J) \sim L^{\Omega} \eqno(39)$$
where $\Omega<1$ is the wandering exponent characteristic to the
sequence[38]. The fluctuating energy per spin then scales as
$\epsilon_{fl}(L) \sim L^{\Omega-1}$. The excess thermal energy per
spin is proportional to the reduced temperature $\epsilon_t(L)\sim t
\sim L^{-1/\nu}$. Comparing $\epsilon_{fl}(L)$ with
$\epsilon_t(L)$ one assumes irrelevant perturbation for
$\epsilon_{fl}(L) \ll \epsilon_t(L)$, i.e. for $\Omega<1-\nu$ and
relevant modulation in the opposite limit $\Omega>1-\nu$. Finally,
the perturbation is marginal for $\Omega=1-\nu$, which condition is
satisfied for the Ising model with $\Omega=0$.

For the hierarchical perturbation considered in this paper the
above criterion can not be used directly. The perturbation is
locally not small and for $r>2$ even
the average value of the coupling $\bar \lambda$ is divergent.
To remedy this problem we consider first the diagonal hierarchy in
eq(2), where the parameters are given as $Y_k=Y_1 r^n$. For a large
$n~~Y_k \sim \exp[2K_d(k)] \sim r^n$, thus the coupling itself is
given as $K_d(k) \sim \log Y_k \sim n K_d(1)$. Therefore
the fluctuating energy in this case is expressed in terms of the
logarithms of the parameters:
$$E_{fl}(L) \simeq \sum_{k=1}^L \log (Y_k/\bar Y) \sim L^{\Omega}
\eqno(40)$$
Similar conclusion is valid for the extreme anisotropic system,
where $K_2(k) \to 0$, $K_1 \to \infty$ and $\lambda_k=K_2(k)
\exp(2K_1)=\lambda_1 r^n$. To that quantum system one can
assign another two-dimensional classical system, if
$K_2$ is kept constant and the vertical couplings $K_1(k)$ vary
with the column index $k$ such that their ratio remains $\lambda_k$.
Then $K_1(k) \sim \log \lambda_k \sim n K_1(1)$, and the
fluctuating energy is again given by:
$$E_{fl}(L) \simeq \sum_{k=1}^L \log (\lambda_k/\bar \lambda) \sim
L^{\Omega}
\eqno(41)$$
Since eqs(40) and (41) remain valid for small aperiodic perturbations
they can be considered as the general definition of the fluctuating
energy both for bounded and unbounded sequences.

For generalized hierarchical sequences the fluctuating energy
can be obtained by noticing that at the critical point
$\prod_{k=1}^L \lambda_k=Z$, which is defined in eq(17). Since $\bar \lambda=
\lim_{L \to \infty}[\prod_{k=1}^L \lambda_k]^{1/L}=1$ we obtain
$$E_{fl}(L)\simeq \log r \left[ {\omega \over m-1}\left({\log L \over
\log m}\right)^{\omega-1} \right] \eqno(42)$$
Thus the wandering exponent $\Omega=0$, for any value of $\omega$
and a linear stability analysis can not decide about the
relevance-irrelevance of the perturbation. In second order of the
analysis the perturbation in the $\omega>1$ region (where $E_{fl}(L)$
logarithmically divergent) is predicted to be relevant, whereas for
$\omega<1$ it is marginally irrelevant. Finally, for
$\omega=1~~E_{fl}(L)$ is independent of the size of the system and
the perturbation is truely marginal. These predictions of the
modified Harris criterion are in complete agreement with our
analytical results.

The critical behavior found at $\omega=1$ is very similar to that
obtained in the two dimensional Ising model with marginally
aperiodic layered interactions. In both type of problems the system
is essentially anisotropic at the critical point and the
corresponding $z$ exponent is related to the surface scaling
dimensions, as in eq(38). The surface spin correlations follow the
scaling law in eq(31), whereas the thermodynamic singularities are
consistent with the anisotropic scaling form of the free-energy
density in eq(35).
Finally, we note that some exponents, including $z,~x_e$ and
$x_{es}$ can be obtained exactly by an analytical method, which can
be used both for aperiodic and hierarchical marginal Ising chains.
Details of the calculations will be presented elsewhere[39].
\vskip 1cm
{\bf Acknowledgement}: This work has been supported by the
Hungarian National Research Fund under grant No OTKA TO12830. F.I.
thanks for interesting discussions with L. Turban, B. Berche and A.
Maritan.
\vfill
\eject
{\bf Appendix}
\bigskip
We calculate the $n_j$ parameter in eq(6) for sequences of lengths
$j=m,m^2,\dots,m^N,\dots$:
$$\eqalign{n_m&=1\cr
n_{m^2}&=n_m m+2^{\omega}-1^{\omega}\cr
&\vdots\cr
n_{m^N}&=n_{m^{N-1}} m+N^{\omega}-(N-1)^{\omega}\cr}$$
which relations are satisfied with
$$n_{m^N}=m^N \left(1-{1 \over m}\right) \sum_{j=1}^{N-1}
{j^{\omega} \over m^j} +N^{\omega}$$
{}From this expression the value of $\delta(\omega,m)$ as given in
eq(9) follows. For a positive integer $\omega~\delta(\omega,m)$ can
be expressed as:
$$\delta(\omega,m)=\left.\left\{(1-x)\left(x{d \over dx}\right)^{\omega}
\left[{1 \over
1-x}\right]\right\}\right|_{x=1/m}$$
The finite size corrections to $n_{m^N}$ are given by:
$$\eqalign{n_{m^N}-\delta(\omega,m) m^N&=- \left(1-{1 \over m}\right)
\sum_{j=N}^{\infty}
{j^{\omega} \over m^j} +N^{\omega}\cr
&=N^{\omega}\left[1-\left(1-{1 \over m}\right) \sum_{j=0}^{\infty}
{(1+j/N)^{\omega} \over m^j}\right]\cr}$$
Expanding $(1+j/N)^{\omega}$ in Taylor series and performing the
summation the leading term is given in eq(16).
\vfill
\eject
{\bf References}
\bigskip
\item{ [1]} D. Schechtman, I. Blech, D. Gratias and J.W. Cahn,
Phys. Rev. Lett. 53, 1951 (1984)
\bigskip
\item{ [2]} C. Godr\'eche, J.M. Luck and H.J. Orland, J. Stat.
Phys. 45, 777 (1986)
\bigskip
\item{ [3]} Y. Okabe and K. Niizeki, J. Phys. Soc. Japan 57, 1536
(1988)
\bigskip
\item{ [4]} E.S. S\/orensen, M.V. Jari\`c and M. Ronchetti, Phys.
Rev. B44, 9271 (1991)
\bigskip
\item{ [5]} S. Sakamoto, F. Yonezawa, K. Aoki, S. Nos\`e and M.
Nori, J. Phys. A22, L705 (1989)
\bigskip
\item{ [6]} C. Zhang and K. De'Bell, Phys. Rev. B47, 8558 (1993)
\bigskip
\item{ [7]} G. Langie and F. Igl\'oi, J. Phys. A25, L487 (1992)
\bigskip
\item{ [8]} B.M. McCoy and T.T. Wu, Phys. Rev. Lett. 21, 549 (1968)
\bigskip
\item{ [9]} C.A. Tracy, J. Phys. A21, L603 (1988)
\bigskip
\item{[10]} F. Igl\'oi, J. Phys. A21, L911 (1988)
\bigskip
\item{[11]} M.M. Doria and I.I. Satija, Phys. Rev. Lett. 60, 444
(1988)
\bigskip
\item{[12]} G.V. Benza, Europhys. Lett. 8, 321 (1989)
\bigskip
\item{[13]} M. Henkel and A. Patk\'os, J. Phys. A25, 5223 (1992)
\bigskip
\item{[14]} Z. Lin and R. Tao, J. Phys. A25, 2483 (1992)
\bigskip
\item{[15]} J.M. Luck, J. Stat. Phys. 72, 417 (1993)
\bigskip
\item{[16]} A.B. Harris, J. Phys. C7, 1671 (1974)
\bigskip
\item{[17]} F. Igl\'oi, J. Phys. A26, L703 (1993)
\bigskip
\item{[18]} L. Turban, F. Igl\'oi and B. Berche, Phys. Rev. B49, 12695
(1994)
\bigskip
\item{[19]} L. Turban, P-E. Berche and B. Berche, J. Phys. A27,
6349 (1994)
\bigskip
\item{[20]} B. Berche, P-E. Berche, M. Henkel, F. Igl\'oi, P.
Lajk\'o, S. Morgan and L. Turban, J. Phys. A28, L165 (1995)
\bigskip
\item{[21]} H.A. Simon and A. Ando, Econometrica 29, 111 (1961)
\bigskip
\item{[22]} B.A. Huberman and M. Kerszberg, J. Phys. A18, L331
(1985)
\bigskip
\item{[23]} S. Teitel and E. Domany, Phys. Rev. Lett. 55, 2176
(1985)
\bigskip
\item{[24]} A. Maritan and A.L. Stella, J. Phys. A19, L269 (1986)
\bigskip
\item{[25]} H. Kunz, R. Livi and A. S\"ut\H o, Comm. Math. Phys.
122, 643 (1989)
\bigskip
\item{[26]} H.A. Ceccatto, Z. Phys. B75, 253 (1989)
\bigskip
\item{[27]} Z. Lin and M. Goda, Phys. Rev. B51, 6093 (1995)
\bigskip
\item{[28]} A.L. Stella, M.R. Swift, J.G. Amar, T.L. Einstein, M.W.
Cole and J.R. Banavar, Phys. Rev. Lett. 23, 3818 (1993)
\item{[29]} A. Giacomtti, A. Maritan and A.L. Stella, Int. J. Mod. Phys.
B5, 709 (1991)
\bigskip
\item{[30]} P. Pfeuty, Phys. Lett. 72A, 245 (1979)
\bigskip
\item{[31]} M.N. Barber, in {\it Phase Transitions and Critical
Phenomena}, edited by C. Domb and J.L. Lebowitz (Academic, London,
1983) Vol.8
\bigskip
\item{[32]} F. Igl\'oi and P. Lajk\'o, Z. Phys. B (in print)
\bigskip
\item{[33]} H.J. Hilhorst and J.M.J. van Leeuwen, Phys. Rev. Lett.
47, 1188 (1981)
\bigskip
\item{[34]} F. Igl\'oi and P. Lajk\'o (unpublished)
\bigskip
\item{[35]} K. Binder abd J.S. Wang, J. Stat. Phys. 55, 87 (1989)
\bigskip
\item{[36]} E.H. Lieb, T.D. Schultz and D.C. Mattis, Ann. Phys. NY.
16, 406 (1961)
\bigskip
\item{[37]} M. Henkel and G. Sch\"utz, J. Phys. A21, 2617 (1988)
\bigskip
\item{[38]} M. Queff\'elec, {\it Substitutional Dynamical
Systems-Spectral Analysis}, Lecture Notes in Mathemathics, Vol 1294
ed. A. Dold and B. Eckmann (Springer, Berlin, 1987)
\bigskip
\item{[39]} F. Igl\'oi (unpublished)
\vfill
\eject
{\bf Figure captions}
\vskip 1cm
\item{Fig.1:} Two-dimensional Ising model with layered hierarchical
couplings: (a) along the col\-umns and (b) along the diagonals.
\vskip 1cm
\item{Fig.2:} Surface magnetisation exponent for marginally
hierarchical Ising chains at the left surface ($\beta_s$) and at the
right boundary ($\bar{\beta_s}$) as a function of the parameter r.
The lines correspond to the analytical results in eqs(13) and (20)
for the (1,0) surface
and the squares are numerical estimates for the (1,1) surface.
\vskip 1cm
\item{Fig.3:} The surface magnetisation for relevant perturbations
($r=.92$) at different values of $\omega$.
\vfill
\eject
{\bf Table caption}
\vskip 1cm
\item{Table 1:} Finite size estimates for the specific heat
exponent of the $m=2$ marginally hierarchical Ising model for
different values of the critical coupling $\lambda_c$. The values
in the table were calculated from finite
size results on chains of length $N,N/4$ and $N/16$. In the last
row the conjectured results from anisotropic scaling eq(37) are
given.
\vfill
\eject
$$\vbox{\settabs 5 \columns
\+N&$\lambda_c^2=2$&$\lambda_c^2=3$&$\lambda_c^2=4$&$\lambda_c^2=5$\cr
\+&&&&\cr
\+32&-.11269&-.27620&-.43191&-.58306\cr
\+64&-.08671&-.23422&-.37389&-.50684\cr
\+128&-.08008&-.21749&-.34636&-.46609\cr
\+256&-.08007&-.21118&-.33350&-.44517\cr
\+512&-.08154&-.20886&-.32748&-.43454\cr
\+1024&-.08288&-.20801&-.32460&-.42922\cr
\+2048&-.08378&-.20771&-.32319&-.42679\cr
\+4096&-.08434&-.20765&-.32285&-.42507\cr
\+&&&&\cr
\+$1-z$&-.08496&-.20752&-.32193&-.42400\cr}$$
\vskip 1cm
\centerline{Table 1}
\vfill
\eject
\end